\documentclass[%
reprint,
showpacs,
longbibliography,
 amsmath,amssymb,
 aps,
]{revtex4-1}

\usepackage{appendix}

\usepackage{graphicx}
\usepackage{dcolumn}
\usepackage{bm}
\usepackage[colorlinks = true,
            linkcolor = blue,
            urlcolor  = blue,
            citecolor = blue,
            anchorcolor = blue]{hyperref}

\usepackage{xcolor}
\usepackage{siunitx}
\usepackage{physics}

\newcommand{\srfc}{$^1$S$_0$\,--\,$^1$P$_1$}

\begin{document}


\title{A Sideband-Enhanced Cold Atomic Source For Optical Clocks}

%
\author{Matteo Barbiero}
\author{Marco G. Tarallo}%
 \email{m.tarallo@inrim.it}
\author{Davide Calonico}
\author{Filippo Levi}
\affiliation{%
 Istituto Nazionale di Ricerca Metrologica, Strada delle Cacce 91, 10135 Torino, Italy}%
\author{G. Lamporesi}
\author{G. Ferrari}


\affiliation{
INO-CNR  BEC  Center  and  Dipartimento  di  Fisica, Universit\`a  di  Trento,  38123  Povo, Italy }%


\date{\today}

\begin{abstract}
We demonstrate the enhancement and optimization of a cold strontium atomic beam from a two-dimensional magneto-optical trap (2D-MOT) transversely loaded from a collimated atomic beam by adding a sideband frequency to the cooling laser. The parameters of the cooling and sideband beams were scanned to achieve the maximum atomic beam flux and compared with Monte Carlo simulations. We obtained a 2.3 times larger, and 4 times brighter, atomic flux than a conventional, single-frequency 2D-MOT, for a given total power of 200 mW. We show that the sideband-enhanced 2D-MOT can reach the loading rate performances of space demanding Zeeman slower-based systems, while it can overcome systematic effects due to thermal beam collisions and hot black-body radiation shift, making it suitable for both transportable and accurate optical lattice clocks. Finally we numerically studied the possible extensions of the sideband-enhanced 2D-MOT to other alkaline-earth species.



\end{abstract}

\maketitle
\section{\label{sec:intro}Introduction}








A cold, bright and compact atomic beam source is 
an important asset for any experiment featuring ultra-cold atoms, such as atom interferometers~\cite{Cronin09}, degenerate quantum gases for quantum simulation~\cite{Georgescu14} and, in particular, optical atomic clocks~\cite{Ludlow2015}. In the last case, two valence electron alkaline-earth (like) metals, like Ca~\cite{Wilpers07}, Sr~\cite{Ushijima2015}, Mg~\cite{Kulosa15}, and Yb\cite{McGrew2018}, are generally used as atomic frequency discriminators, and present low vapor pressures at room temperature thus needing high temperature ovens to generate enough atomic vapour, typically followed by a space-demanding Zeeman slower (ZS). Although compact and transportable versions of the ``oven + ZS'' atomic beam system have been developed~\cite{Poli2014,AOsense}, some concerns about the systematic effects due to collisions with the atomic beam particles~\cite{Gibble13} and the hot black-body radiation from the oven region~\cite{Beloy14} can arise below the $10^{-18}$ relative uncertainty level.

The two-dimensional magneto optical trap (2D-MOT) atomic source~\cite{Dieckmann1998, Schoser2002} can be transversely loaded, hence reducing the setup dimensions, avoiding direct exposure of the atomic reference to hot metals, and at the same time obtaining an optical shutter of the atomic beam just by turning-off its cooling beams. This avoids the use of in-vacuum mechanical shutters or optical beam deflectors~\cite{Witte92} as done for ZS or collimated oven beams. The 2D-MOT system complexity can be further reduced by its permanent magnets implementation~\cite{Tiecke2009, Lamporesi2013}.


In this work, we present a novel atomic source employing a 2D-MOT source of strontium (Sr) atoms for metrological application. The mechanical implementation of the atomic source is similar to other setups built to generate lithium~\cite{Tiecke2009}, sodium \cite{Lamporesi2013,Colzi2018} and strontium~\cite{Nosske2017} atomic beams. Our system is further characterized by a collimated atomic beam transmitted by a bundle of capillaries directly towards the 2D-MOT region, and a two-frequency optical molasses to enhance the atomic flux toward the trapping region. The design, engineering and characterization of the sideband-enhanced 2D-MOT strontium source is the main result of this work. This is accomplished by looking at the loading performances of a three-dimensional MOT typically used as the first cooling and trapping stage for an optical lattice clock~\cite{Xu2003}. Monte Carlo (MC) numerical simulations are used to find the optimal optical configuration which are then compared to the experimental results. 

The article is organized as follows: 
Sec.~\ref{sec:Theory} introduces the physical interpretation and significance of adding a sideband frequency to the cooling beams of the 2D-MOT; Sec.~\ref{sec:apparatus} depicts the experimental apparatus assembled for an optical lattice clock; in Sec.~\ref{sec:numerical simulation} we describe the numerical modeling of the atomic source and the 2D-MOT cooling and trapping processes by Monte Carlo simulations; Sec.~\ref{sec:atomic source characterization} shows the experimental characterization of our atomic source and in Sec.~\ref{sec:sideband enhancement} we demonstrate how the sideband-enhancement method is able to magnify the number of trapped atoms by a magneto-optical trap.


\section{\label{sec:Theory}Principles of sideband-enhanced 2D-MOT}

\begin{figure}[tb]
    \centering
    \includegraphics[width=0.49\textwidth]{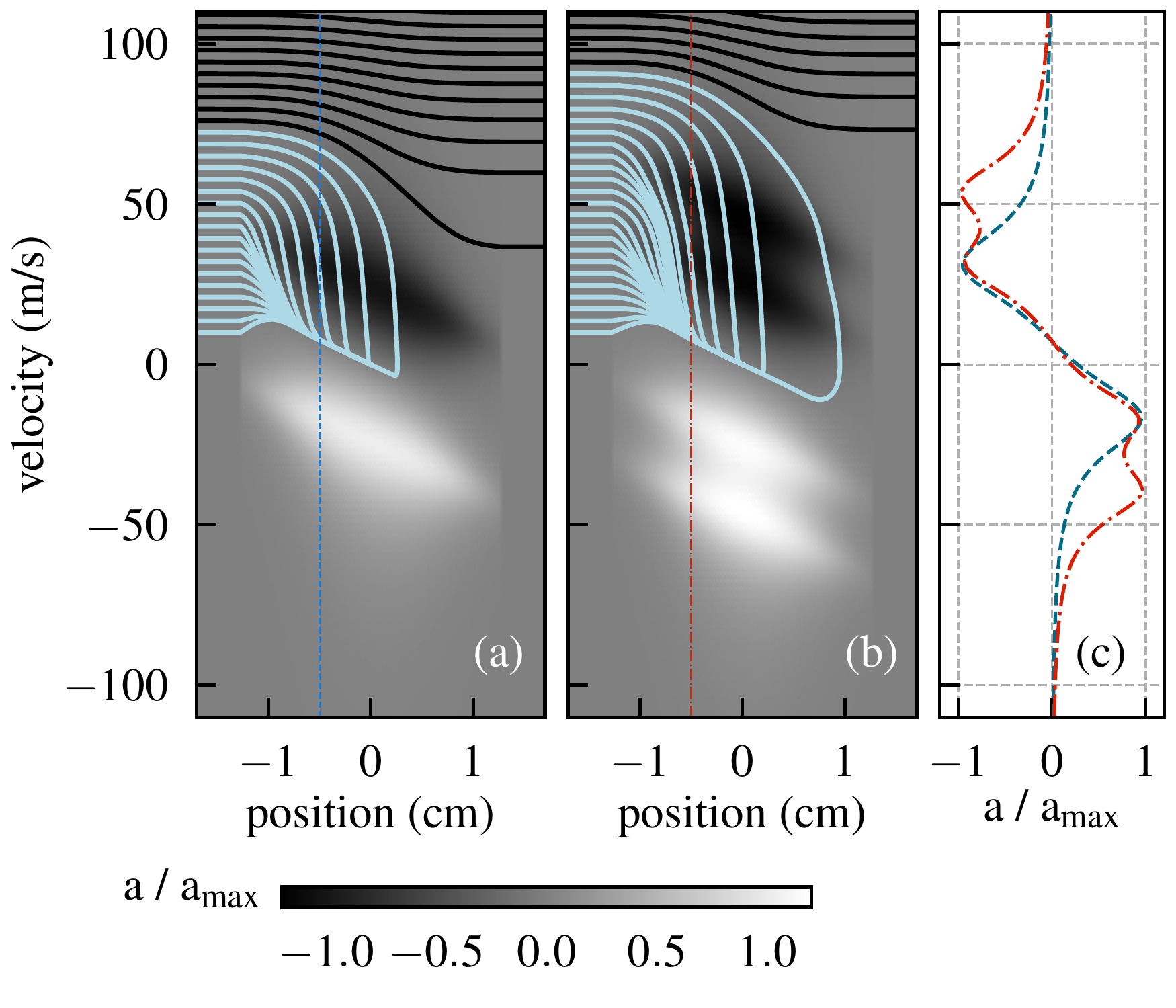}
    \caption{1D simulation of the atomic trajectories of strontium atoms (light blue line) for different capture processes in a 2D-MOT. The color map plotted on the background depicts the acceleration value at each point of the phase-space. \textbf{(a)} Single-frequency 2D-MOT. The total saturation and the detuning of the MOT beams are $s=7$, $\Delta/\Gamma = -1.6$. The estimated capture velocity is \SI{72(1)}{m \per s}. 
    \textbf{(b)} Sideband-enhanced 2D-MOT. The saturation parameter is $s=3.5$ for the 2D-MOT beam at with $\Delta/\Gamma =-1.6 $, and $s_\text{SB}=3.5$ for the sideband beam  at $\Delta_\text{side}/ \Gamma = -3.2$. The estimated capture velocity is \SI{90(1)}{m \per s}. The magnetic field gradient and beam width used in this calculation are $b=\SI{0.22}{T \per m}$ and $w=\SI{1}{cm}$.
    \textbf{(c)} Acceleration profile at $r=-w/2$ of the sideband trapping (red line) and  standard 2D-MOT  trapping (blue line).}
    \label{fig:capt_vel}
\end{figure}

A 2D-MOT atomic source relies on the radiation-pressure friction force to capture and cool thermal atoms effusing from either an oven, or a background gas. In this work, we focus our attention on the 2D-MOT loaded from a collimated atomic source, so that a 1D model offers a good insight on the expected 2D-MOT flux. For the 1D model, the MOT captured atoms per second  $\Phi_\text{2D}$ is provided by the formula 
\begin{equation}\label{eq:cap_2d}
\Phi_\text{2D} \simeq n v_\text{th} A (v_\text{c} / v_\text{th})^4, \qquad(v_c \ll v_\text{th})
\end{equation}
where $n$ is the spatial density of the thermal beam, $v_\text{th} =  \sqrt{2 k_B T_\text{ov} / m}$ is the most probable thermal velocity (for the atomic Sr vapour at $T_\text{ov} = \SI{460}{\celsius}$, $v_\text{th} = \SI{379}{m \per s}$ ), $A = 2\pi w^2$ is the MOT capture surface related to the trapping beam width $w$, and $v_c$ is the capture velocity of the trap. It is clear from (\ref{eq:cap_2d}) that the most influential parameter is the capture velocity $v_\text{c}$, which is related to the magnetic gradient $b$, the frequency detuning $\Delta$ from the cooling transition, and the total saturation parameter $s=I/I_\text{sat}$ of the MOT optical beams, where for an atomic transition at wavelength $\lambda$ and spontaneous emission rate $\Gamma$ the resonant saturation intensity is $I_\text{sat} = \pi hc\Gamma/3\lambda^3$. In the 1D model one typically computes $v_\text{c}$ numerically by solving the semiclassical equation of motion, as shown in Fig.\ref{fig:capt_vel}(a). Here one can observe that there are two different dynamics inside the MOT region. In the outer region, the MOT behaves like a Zeeman slower, where the friction force exerted upon any atom will be effective only if the velocity $v$ at distance $r$ from the symmetry axis will be nearly resonant with the cooling laser, i.e., if the difference of the Zeeman shift and the laser detuning equals the Doppler shift. In the inner region the motion of the atoms can be described by an overdamped harmonic oscillator model. Hence the capture velocity is strictly related with the dynamics in the outer region of the MOT and, assuming perfect compensation of the Zeeman shift and Doppler shift, it can be roughly estimated as~\cite{Tiecke2009,Zinner98}

\begin{equation}\label{eq:vc}
v_\text{c} \lesssim v_\text{max} = \sqrt{a_\text{max}\,r_\text{max}}    
\end{equation}
where $a_\text{max} = \hbar k \Gamma/(2m)$ is the maximum acceleration at infinite saturation parameter, and $r_\text{max}=\sqrt{2}w$ is the maximum interaction distance with the MOT beams, taking into account the projection of the 2D-MOT beams at \SI{45}{\degree} from the atoms propagation axis. This $v_\text{c}$  corresponds to the maximum velocity allowed in order to decelerate an atom to zero at the center of the trap. This oversimplified estimation gives us some hints on the 2D-MOT expected performance. In particular, even for infinite available power, the capture velocity would be bounded, while the capture mechanism is fundamentally limited by the natural linewidth of the cooling transition and the cooling beam radius. However, if one uses laser light which has several red-detuned sidebands, even faster atoms can be slowed down and the capture velocity increased. MOT loading enhancement was observed in alkali atomic systems by means of electro-optic modulation (EOM) of the cooling beams \cite{Anderson1994,Lee2017}. This technique is generally not feasible at the wavelengths of alkaline-earth atoms by EOMs. Furthermore, because of the higher $\Gamma$s excessive spectral broadening would reduce the radiation pressure force, making it no longer sufficient to keep the thermal atoms in the trap. A one-sideband 3D MOT has been previously realized to trap Ca atoms loaded directly from an effusing atomic oven~\cite{Zinner98,Riehle:1999p6441}. In this case, with a total MOT saturation parameter $s \sim 0.1$ and an atomic vapour temperature of 600 $^\circ$C, an enhancement factor of 7 was observed~\cite{Riehle:1999p6441}. However here only a very small fraction of the available atoms were trapped, hence that system would be very unfavourable in the case of a 2D-MOT source loading. 

It is more interesting to investigate the sideband-enhanced 2D-MOT in the limit of high total saturation parameter $s\geq 1$, where most of the low velocity class ($v\leq v_\text{max}$) is slowed and captured by the cooling beams. Fig.\ref{fig:capt_vel}(a) shows a simulation of the phase-space trajectories for typical values of the experimental parameters ($\Delta,s,b$) used in a strontium 2D-MOT~\cite{Nosske2017}. The acceleration patterns of the sideband-enhanced 2D-MOT in the atomic phase-space are depicted in the in Fig.\ref{fig:capt_vel}(b). As shown in the plot, the sideband beams interact with atoms from a higher velocity class, decelerating them toward the capture region of the standard MOT beam. This increment of the capture velocity is best displayed in Fig.\ref{fig:capt_vel}(c): here we can see the MOT acceleration as function of the atomic approaching velocity. In the standard MOT (blue dashed) the force is peaked around a given velocity value, reaching $a_\text{max}$ and the amount of power increases the spectral width of the force as $\sqrt{s}$. On the other hand, the sideband-enhanced force (red dot-dashed) presents a second peak at higher velocity without degrading the peak acceleration. Optimal positioning of the sideband frequency thus allows an increase of the expected capture velocity $v_c$ and of the expected MOT loading rate too. 

Another expected beneficial effect of the sideband-enhanced 2D-MOT with large $s$ is the reduction of the transverse temperature of the cold atomic sample compared to the standard 2D-MOT, which would yield a higher brightness (i.e. lower beam divergence). This can be explained considering that the optical power redistributed at a higher frequency weakly interacts with the atoms trapped once they reach the center of the MOT.

In order to correctly address the expected performances of a sideband-enhanced strontium 2D-MOT we performed a dedicated Monte Carlo (MC) simulation which takes into account the actual geometry of the system, the magnetic field gradient, the residual divergence of our atomic beam from the oven, and the expected loading rate for the final 3D-MOT. This is described in detail in Sec.\ref{sec:numerical simulation}.

\section{\label{sec:apparatus}Experimental apparatus}

\subsection{Vacuum system}
\begin{figure}
    \centering
    \includegraphics[width=0.49\textwidth]{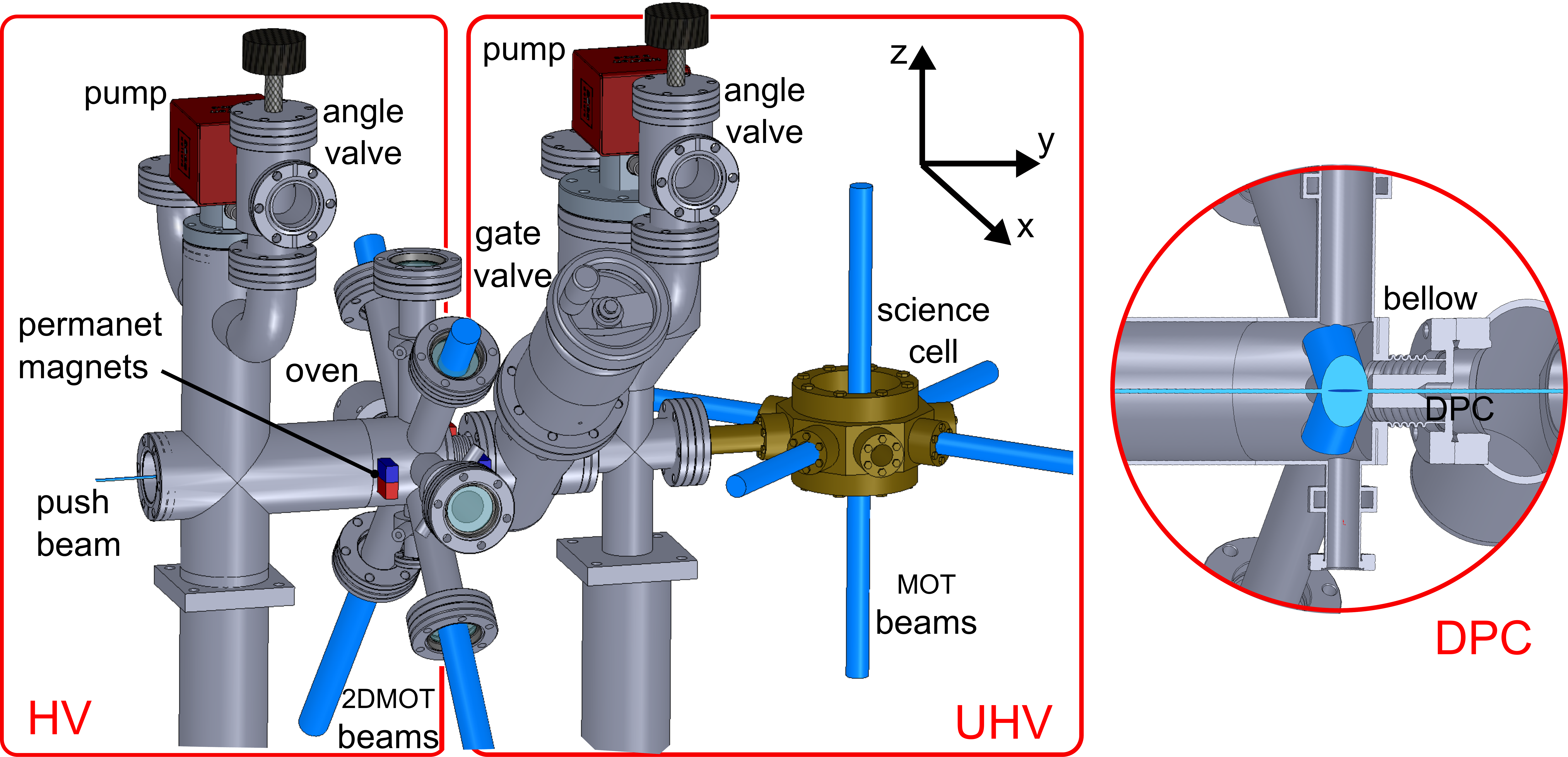}
    \caption{Schematic drawing of the vacuum system. It hosts a high vacuum (HV) region for the atomic beam production, and an ultra-high vacuum (UHV) region for cooling and trapping the atomic sample. A DPC connects the two regions. The size of the entire vacuum apparatus is roughly \SI{70}{cm} $\times$ \SI{70}{cm} $\times$ \SI{45}{cm}.}
    \label{fig:Figure1}
\end{figure}

The schematic drawing of the vacuum system to produce and trap ultra-cold strontium atoms is depicted in Fig.~\ref{fig:Figure1}. It has been previously described in~\cite{Tarallo2017}, and its concept is adapted from previous works~\cite{Tiecke2009, Lamporesi2013}. The vacuum system is conceived to host two physical regions with very different vacuum levels, the atomic source region and the science cell region and, at the same time, to be very compact. The atomic source region consists of a stainless-steel vacuum chamber with a multi-way cross  at its end, where the intersection plane of the tubes forms the 2D-MOT plane. The ultra-high vacuum region hosts a small octagonal science cell with two large vertical optical accesses (DN63CF) and seven small lateral optical windows (DN16CF) for cooling, trapping and operate a Sr optical clock. The two vacuum regions are connected by a differential pumping channel (DPC) carved in a custom bellow with \SI{2}{mm} diameter and \SI{22.8}{mm} length and all-metal gate valve. The DPC sets the maximum divergence of the cold atomic beam at \SI{87}{mrad}, while a conductance of \SI{4.3e-2}{L \per s} allows to maintain a differential pressure of \num{e4} between the two regions. Vacuum is maintained by two ion-getter pumps, both regions reaching a pressure below \SI{e-10}{mbar} when the oven is not heated.


%

\subsection{Collimated atomic source}

The oven consists of a simple stainless-steel cylinder with an aperture of \SI{16}{mm} and a conflat flange DN16CF to be attached to the main body of the vacuum system on one of its circular sides.
The oven is attached to the multi-way cross vacuum chamber \SI{128}{mm} away from its center.
In order to produce a collimated atomic beam, an array of $N_\text{cap} \simeq \num{150}$ capillaries made of nickel-based alloy Monel400, with an internal radius $r_\text{cap}=\SI{0.2}{mm}$ and a length  $L_\text{cap}=\SI{20}{mm}$, is inserted at the oven aperture. The capillaries are tightened inside a holder which lays in the aperture of the oven. The heating is insured by a pair of heating cartridges. The Sr vapor is typically generated at the temperature $T_\text{ov} =\SI{460}{\celsius}$. In order to avoid clogging of the capillaries with strontium, the oven hosts an extra pair of heating cartridges close to its aperture to maintain the capillaries at a temperature $T_\text{cap}$ higher than $T_\text{ov}$. For all experimental characterizations we maintained a differential temperature $T_\text{cap} - T_\text{ov}=\SI{30}{\celsius}$. At the typical operational oven temperature $T_\text{ov}=\SI{460}{\celsius}$, the estimated vapour pressure inside is $p_\text{ov}=\SI{0.133}{Pa}$~\cite{Alcock1984} from which we estimate the Sr atomic density by means of the ideal gas law $n_\text{ov}= p_\text{ov} / k_B T_\text{ov} =\SI{1.31e19}{atoms \per m^3}$.
In the regime of negligible collisions inside the capillaries (mean free path $\lambda_\text{ov} = (\sqrt{2} n_\text{ov} \sigma_\text{Sr} )^{-1}\sim \SI{70}{mm} \gg L_\text{cap}$, with $\sigma_\text{Sr}$ = 8$\cdot10^{-19}$ m$^2$ the elastic cross section), the atomic flux is proportional to the oven pressure $P_\text{ov}$  and it is estimated as~\cite{Wang1960}:
\begin{equation}
    \Phi_\text{ov} =a \frac{4 \sqrt{\pi}}{3} \frac{n_\text{ov} v_\text{th} r^3_\text{cap}}{L_\text{cap}}N_\text{cap}
    \label{eq:flux_oven}
\end{equation}
where $a$ is the isotopic abundance.
In the case of $^{88}$Sr, the expected atomic flux  at $T_\text{ov}= \SI{460}{\celsius}$ is $\Phi_\text{ov}=\SI{5.8e14}{atoms \per s}$. The geometrical constraint imposed by the capillaries yield a theoretical divergence angle $\theta_\text{cap} \simeq r_\text{cap} / L_\text{cap}=20\,$mrad.

\subsection{2D-MOT and cold atomic source generation}

As sketched in Fig.~\ref{fig:Figure1}, the 2D MOT is composed of a 2D quadrupole magnetic field in combination with two orthogonal pairs of retroreflected laser beams of opposite circular polarization.

The magnetic field gradient is generated by four stacks of permanent magnets~\cite{Tiecke2009}. Each stack is composed of \num{9} neodymium bar magnets with size of \SI{25}{mm} $\times$ \SI{10}{mm} $\times$ \SI{3}{mm} and magnetization \SI{6.6(1)e5}{A \per m}. The stacks are placed around the center of the 2D-MOT at the positions $\mathbf{r}_\text{m} = \pm \mathbf{x}_0  \pm \mathbf{y}_0$ where $x_0 = \SI{110}{mm}$ and $y_0 = \SI{90}{mm}$. 
The magnetization of each permanent magnet has been oriented in such a way that it has the same direction with the one along the $y$-axis and opposite direction with the one faced along the $x$ axis. We estimated the generated field upon the 2D-MOT plane by finite element analysis (FEA). This shows a uniform linear gradient $\mathbf{B}_\text{m} (\mathbf{r}) = b \mathbf{x }  - b \mathbf{z}$ close to  the center of the trap $|\mathbf{r}|<\SI{1}{cm}$ with $b=\SI{0.224}{T \per m}$. As compared to the above expression the maximum deviation of the actual magnetic field is negligible within the 2D-MOT trapping volume, as it ultimately amounts to $\Delta_\text{Z} = 2 \pi\times \SI{5.6}{MHz} = \SI{0.17}{\Gamma}$ in frequency detuning.


The two pairs of counterpropagating beams allow magneto-optical cooling and trapping of slow atoms effusing from the oven along the $x$ and $z$ axes, while they are free to drift along the $y$ direction. Hence a nearly-resonant laser ``push'' beam is directed to the 2D-MOT center along the $y$-axis toward the UHV science cell in order to launch atoms collected in the 2D-MOT towards the MOT region. The center of the MOT in the science cell is located \SI{370}{mm} far from the 2D-MOT center. Finally, the mandatory MOT quadrupolar field is generated by a pair of coils with the current flowing in the Anti-Helmholtz configuration, which generates a typical magnetic field gradient of \SI{0.4}{T \per m}.

\subsection{Laser system}

\begin{figure}[tb]
    \centering
    \includegraphics[width=0.4\textwidth]{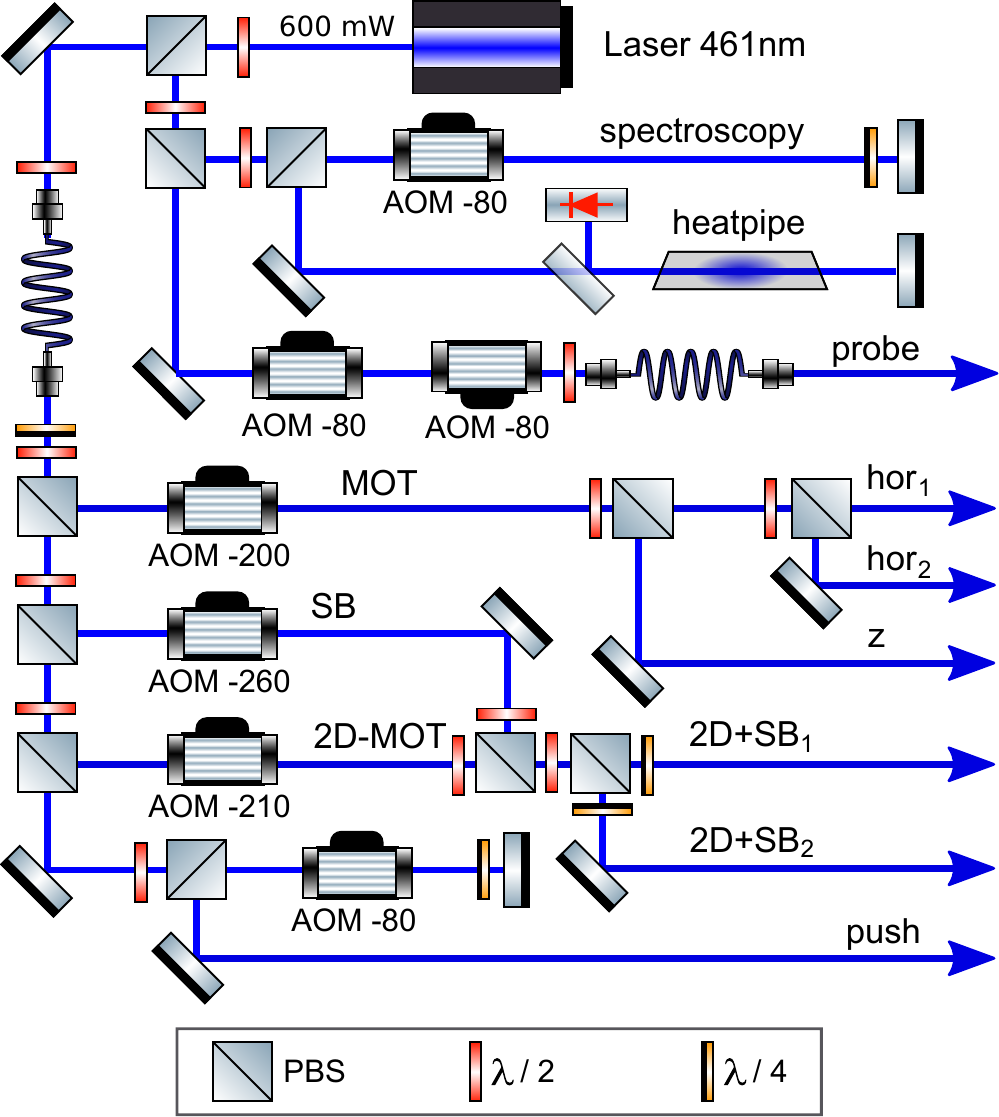}
    \caption{Optical setup of the blue laser system used for cooling, trapping and probe ultracold strontium atoms. Each acousto-optical modulator (AOM) is drawn with its driving frequency (MHz) and the sign corresponds to the diffraction order. Beam shaping lenses are not shown.}
    \label{fig:blue_optical_setup}
\end{figure}

A schematic drawing of the laser system is shown in Fig.~\ref{fig:blue_optical_setup}.
The \SI{461}{nm} laser is provided by a semiconductor-based commercial laser composed of an infrared master laser, a tapered amplifier and a second harmonic generation cavity. It is able to generate up to \SI{600}{mW} of blue power. This blue laser is split in six main optical paths and frequency manipulated by acousto-optic modulators (AOMs). The laser frequency is stabilized to the Sr atomic transition \srfc~by performing wavelength modulation saturation spectroscopy on an hot vapour of strontium generated in a heatpipe~\cite{Poli2006}. 
Typically we are able to deliver about half of the available power to the atoms.

A detailed scheme of the various beam paths is depicted in Fig.\ref{fig:blue_optical_setup}. For typical experimental conditions, the 2D-MOT and sideband beams share \SI{200}{mW} and have a $1/e^2$ beam width $w_\text{2D} =$~\SI{9.5}{mm}, the MOT beams have a total power of \SI{45}{mW} with a beam width of \SI{6.2}{mm} and a detuning from the atomic resonance of -1.2 $\Gamma$, the push beam has a power up to \SI{5}{mW} and a beam width of  \SI{0.81}{mm}, the spectroscopy beam sent inside the heatpipe has a power of \SI{0.5}{mW} and beam width \SI{0.37}{mm}, and finally the probe beam has \SI{0.5}{mW} of power and \SI{0.83}{mm} width. The detuning from the atomic resonance of the beams used in atomic source system (2D-MOT, sideband and push beams) have been scanned for optimal atomic flux, as described in Sec.\ref{sec:atomic source characterization}.


We generate the 2D-MOT main and sideband beams as follows: two dedicated \SI{200}{MHz} and \SI{350}{MHz} AOMs are employed to shift the frequencies of two beams, which are shaped with the same telescope in order to have the same beam width. They are combined in a polarizing beam splitter (PBS) cube with orthogonal linear polarizations in such a way that the 2D-MOT (sideband) beam is completely transmitted (reflected). The two beam polarization is then rotated 45$^{\circ}$ by a half-wavelength retarding waveplate, thus the two beams are recombined into a second PBS which yields the two beams for the branches of the 2D optical molasses. The \SI{350}{MHz} is dedicated to the sideband beam, offering a \SI{100}{MHz} bandwidth to find the optimal frequency which maximizes the loading of the atomic source.

\section{Numerical simulation of the 2D-MOT\label{sec:numerical simulation}}


Monte Carlo (MC) simulation is a powerful and versatile numerical approach because it allows to study complex physical processes in a realistic environment: from the simple MOT capture process \cite{Wohlleben2001,Kohel2003,Chaudhuri2006,Szulc2016}, to the loading process of an optical potential \cite{Hanley2017,Mu2010}, a molecular MOT \cite{Comparat2014} and Rydberg-dressed MOT \cite{Bounds2018}. Knowing the atom-light interactions and the geometry of the system, we want to extract the capture efficiency of our 2D-MOT system at a given trapping configuration, defined by the 2D-MOT beams, sideband beams and push beam, as described in the previous section. The MC algorithm is implemented in Python language.

 We simulate $N_\text{sim}=2 \times 10^4$ trajectories of atoms that interact with the trap. At initial time $t=0$ the starting positions of atoms are randomly sampled in a disk region of radius $r_0 =\SI{7.5}{mm}$ in the $(y,z)$ plane and at $x_0=\SI{-128}{mm}$ far from 2D-MOT trap center along the direction of hot atomic flux emitted by the oven. The velocity space is sampled from the Maxwell-Boltzmann probability distribution expressed in polar coordinates. The sampling of the absolute value of the starting atomic velocity $v_0$ is limited to $v_\text{cut} =\SI{90}{m \per s}$ to speed up the calculation. The polar angle $\theta_0$ is uniformly sampled considering the geometrical constraint imposed by the capillaries $\theta_0 \leq \theta_\text{cap}$. The azimuthal angle $\phi_0$ is randomly chosen between 0 and $2 \pi$.

The trajectory is discretized in time with a step size $\delta t = \SI{50}{\micro s}$, and computed until $t_\text{tot}=\SI{4}{ms}$ by using a Runge-Kutta algoritm~\cite{Enright1989}. The time step $\delta t$ is chosen to be greater than the internal atomic time scale $\tau_{^1P_1} = \Gamma^{-1}$, so that the atom-light interaction can be calculated by using the semi-classical approximation of the  Optical Bloch Equations, but shorter than the capture time for an atom moving at $v_\text{max}$ which is about $\Delta t_\text{max}$ = 165 $\mu$s. At each time step $t_i=i \delta t$, the atom-light scattering rate with a single laser beam is computed as:
\begin{equation}
    R(t_i) = \frac{\Gamma}{2} \frac{s(\mathbf{r}(t_i))}{1 + s(\mathbf{r}(t_i)) + 4\left(\frac{ \Delta_\text{eff}(\mathbf{r}(t_i),\mathbf{v}(t_i))}{ \Gamma}\right)^2}
    \label{eq:scattergin_rate}
\end{equation}
where $s(\mathbf{r}(t_i))$ is the position-dependent saturation parameter, and $\Delta(\mathbf{r}(t_i),\mathbf{v}(t_i))$ is the frequency detuning due to the Doppler and Zeeman  shift. 
The local saturation parameter is computed as:
\begin{equation}
\qquad s(\mathbf{r}(t_i)) = s_0 \exp \left( -\frac{ 2 |\mathbf{r}(t_i) \times \mathbf{\hat{k}}|^2}{ w^2}\right),
\end{equation}
where $s_0$ is the saturation peak, $w$ is the width of the optical beam and where the vector product $\mathbf{r} \times \mathbf{\hat{k}}$ is the distance between the atom position and the center of the laser line propagation described by the unitary vector $\mathbf{\hat{k}}$. Considering the aperture of the optics elements, a spatial cut-off of $|\mathbf{r} \times \mathbf{\hat{k}}|< \SI{1.2}{cm}$ in the local saturation parameter is also applied.
The frequency detuning is computed as:
\begin{equation}
    \Delta_\text{eff}(\mathbf{r}(t_i),\mathbf{v}(t_i)) = \Delta + \mathbf{k}\cdot \mathbf{v}(t_i) -\frac{\mu_B}{\hbar}| \mathbf{B}(\mathbf{r(t_i}))|
\end{equation}
where $\Delta$ is the laser frequency detuning from the atomic transition, $\mathbf{k}\cdot \mathbf{v}(t_i)$ is the Doppler shift and the last term is the Zeeman shift induced by the atomic position in the magnetic field $\mathbf{B}(\mathbf{r}(t_i))$ described in Sec.\ref{sec:apparatus}.

\begin{figure}[t]
    \centering
    \includegraphics[width=0.49\textwidth]{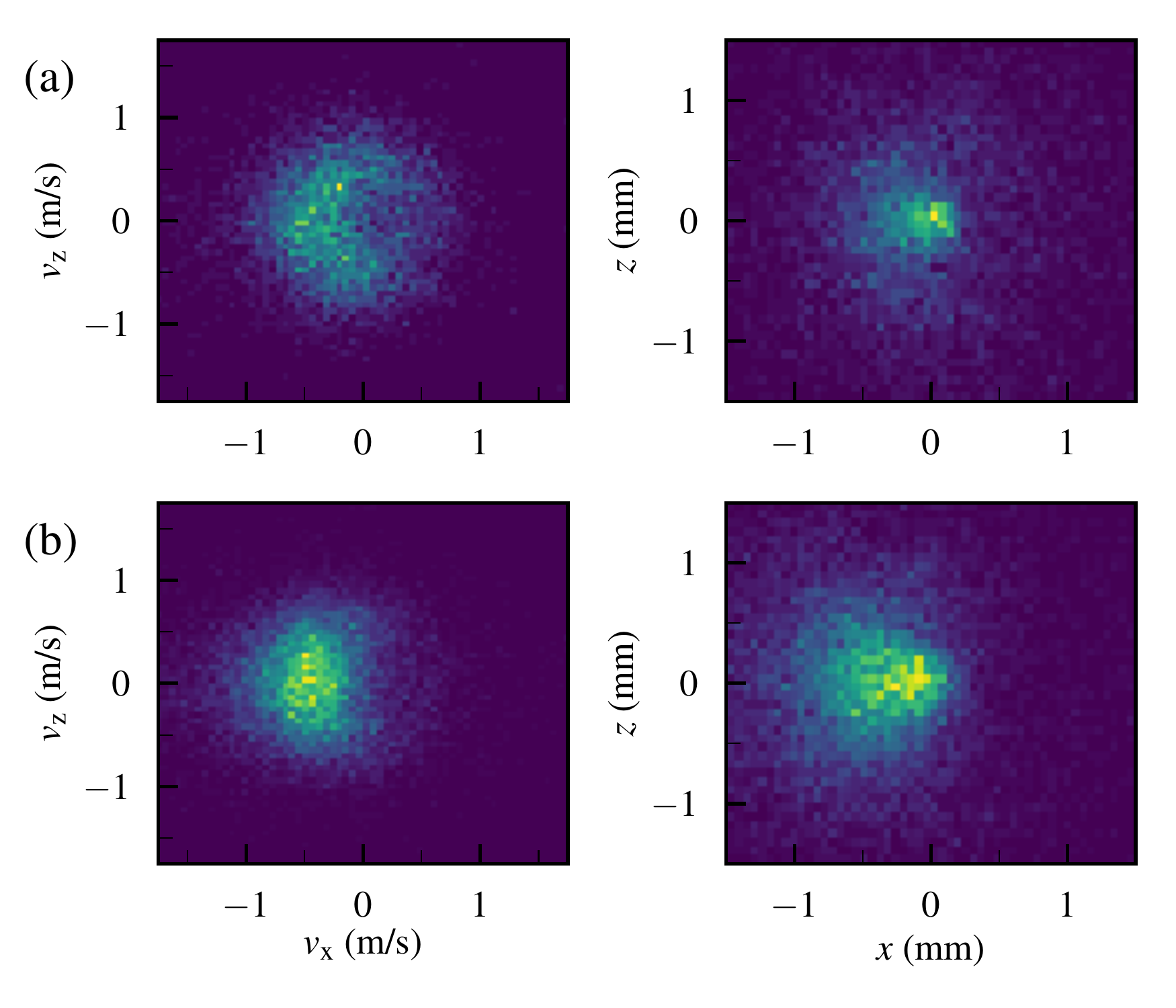}
    \caption{Monte Carlo simulation of the velocity and position transverse coordinates of the 2D-MOT generated atomic beam at the end of the simulation time $t_\text{tot}$. Density maps of the captured trajectories for (a) the single frequency 2D-MOT ($s_\text{2D}$ = $s_{tot}$ = 6.56, $\Delta_\text{2D}$ = -1.6 $\Gamma$), and (b) adding a sideband beam ($s_{SB}$ = 3.45, $s_{2D} = s_{tot} -s_{SB}$, $\Delta_{SB}$ = -3.1 $\Gamma$).}
    \label{fig:mc_sim_hist}
\end{figure}

The heating induced by the spontaneous emission process is also taken into consideration in the simulated dynamics by adding a random recoil momentum $\hbar |k| \sqrt{R \delta t} \ \hat{\mathbf{e}}$, where ${R\delta t}$ is the average number of scattering events in a time interval $\delta t$, while $\mathbf{\hat{e}}$ is a unitary vector randomly chosen from an isotropic distribution~\cite{Kohel2003}. The resulting atom's acceleration induced by the 2D-MOT (and sideband) beams is described according to
\begin{equation}
    \mathbf{a}_\text{2D,SB}= \frac{\hbar |k| }{m} \sum_{n=0}^{4} \frac{R_n}{4} \left[   \hat{\mathbf{k}}_n + \frac{\hat{\mathbf{e}}_n}{\sqrt{R_n \delta t / 4}}\right], 
\end{equation}
where the saturation peak $s_\text{2D,SB}$ is redistributed equally among the 4 beams of the 2D-MOT and sideband and  the beam directions $\mathbf{\hat{k}}_n$ are described by the 4 combinations of the unitary vectors $(\pm \mathbf{\hat{x}} \pm \mathbf{\hat{z}}) / \sqrt{2}$.
The acceleration induced by the push beam  is computed as:
\begin{equation}
    \mathbf{a}_\text{push}= \frac{\hbar |k| }{m} R_\text{push} \left[   \hat{\mathbf{y}} + \frac{\hat{\mathbf{e}}}{\sqrt{R_\text{push} \delta t }}\right] .
\end{equation}

The total acceleration $\mathbf{a}(t_i)$ exerted on the atom at position $\mathbf{r}(t_i)$ with velocity $\mathbf{v}(t_i)$  is quantified as the sum of the above processes:
\begin{equation}
    \mathbf{a} =  \mathbf{a}_\text{2D}  + 
     \mathbf{a}_\text{SB}  +
     \mathbf{a}_\text{push}  
\end{equation}


Once $t=t_\text{tot}$, each simulated atom is considered captured in the MOT if the divergence of the atomic trajectory computed along the push direction is lower that the geometrical constraint imposed by the MOT capture angle $\theta_\text{MOT} = \SI{16}{\milli rad}$ and if the final longitudinal velocity is below the MOT capture velocity  $v^\text{MC}_\text{capt} =\SI{60}{m \per s}$. In the selection of the captured trajectories, we also considered the losses due to collisions with hot atoms from the thermal beam, whose time scale is calculated to be $\tau_\text{coll}=\SI{50}{ms}$. Hence for each trajectory the collision probability is estimated as $p_\text{coll} = 1-e^{- \tau_\text{2D} / \tau_\text{coll}}$ and a unitary random number $\varepsilon$ is generated in order to accept ($\varepsilon>p_\text{coll}$) or reject ($\varepsilon<p_\text{coll}$) each simulated atomic trajectory. Figure~\ref{fig:mc_sim_hist} shows the results from two simulation runs, where the final transverse velocity and position coordinates are displayed versus the number of occurrences. The resulting velocity distribution is used to estimate the transverse temperature of the atomic beam.

Finally, a capture efficiency ratio $r$ is defined as $r = N_\text{tr} / N_\text{sim}$, where $N_\text{tr}$ is the number of captured trajectories for a given trapping configuration. Besides a scaling factor, the $r$ parameter is used as a comparison with the experimental data. The numerical results of such modeled atomic source are presented in Sec.\ref{sec:atomic source characterization} and \ref{sec:sideband enhancement} together with the experimental data.



\section{Atomic source characterization \label{sec:atomic source characterization}}

\begin{figure*}[t]
    \centering
    \includegraphics[width=\textwidth]{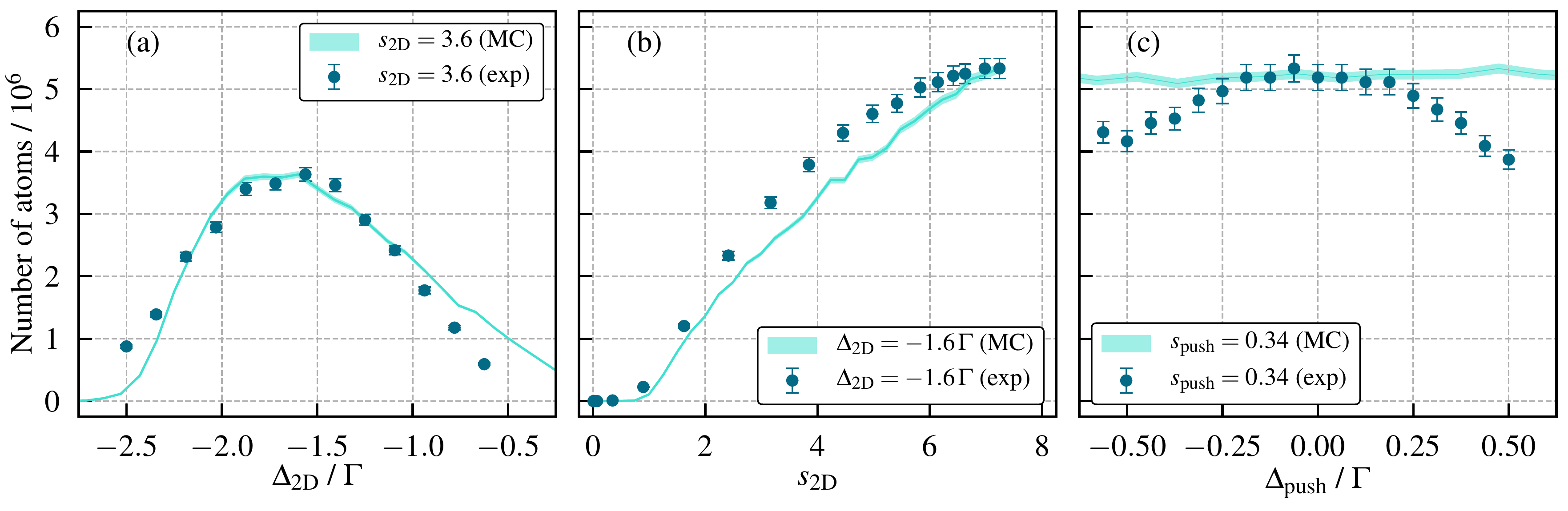}
    \caption{Experimental characterization of the strontium atomic source in the single-frequency 2D-MOT configuration. \textbf{(a)} Number of atoms loaded in the MOT as a function of the 2D-MOT detuning $\Delta_\text{2D}$ at fixed saturation parameter $s_\text{2D}=3.6$. The blue points are the experimental values of the number of atoms in the MOT. The turquoise regions are the MC simulations properly scaled with the experimental points. \textbf{(b)} Number of atoms in the MOT as function of the 2D-MOT saturation $s_\text{2D}$ at fixed detuning $\Delta_\text{2D}= - 1.6 \, \Gamma$.
    \textbf{(c)} Number of atoms in the MOT as a function of the push detuning $\Delta_\text{push}$ at fixed saturation $s_\text{push}$= 0.34. All the data were taken at $T_\text{ov}=\SI{460}{\celsius}$.}
    \label{fig:2DMOT_char}
\end{figure*}



In Fig.~\ref{fig:2DMOT_char} we report the results of the characterization of our strontium 2D-MOT atomic source. This was obtained by looking at the loading of the MOT in the science chamber. Laser parameters of the 2D-MOT and push beams,  $\Delta_\text{2D}$, $s_\text{2D}$ and $\Delta_\text{push}$, were scanned for optimal settings in order to find the maximum loading rate (blue points) and then compared with the expected capture ratio $r$ from MC simulation (turquoise region). 

Regarding the 2D-MOT beam parameters, Fig.\ref{fig:2DMOT_char}(a) shows the number of atoms loaded in the MOT that reaches its maximum value at $\Delta_\text{2D} = -1.6\, \Gamma$, with a FWHM of the order of $1.5\,\Gamma$. The peak position and the spectral response are in good agreement with the simulated one. Fig.\ref{fig:2DMOT_char}(b) shows the increase of number of atoms in the MOT as a function of the 2D-MOT optical intensity at $\Delta_\text{2D}=-1.6\,\Gamma$. Here we can observe that for $s_\text{2D}>6$, the number of atoms in the MOT starts to saturate, however much later than the unity value. The same result is predicted by the MC simulation. 

We observed an optimal push intensity around the $s_\text{push} \simeq  0.34$, beyond this value the MOT number of atoms decreases, as previously verified in a similar setup~\cite{Nosske2017}. The reduced efficiency in the transfer from the 2D-MOT to the blue MOT is explained considering that atoms accelerate beyond $v_\text{capt}$ cannot be captured in the MOT. This behaviour at higher $s_\text{push}$ is  also observed in the MC simulation considering only the atom captured in the MOT at longitudinal $v_L$ velocity below $v^\text{MC}_\text{capt} \sim \SI{60}{ \per s}$. Fig.\ref{fig:2DMOT_char}(c) shows the MOT number of atoms as a function of the push beam detuning. From this plot we observe that the best transfer efficiency is obtained near the atomic resonance $\Delta_\text{push} = 0$, but it is not a critical parameter.

\begin{figure}[!b]
 \includegraphics[width=0.45\textwidth]{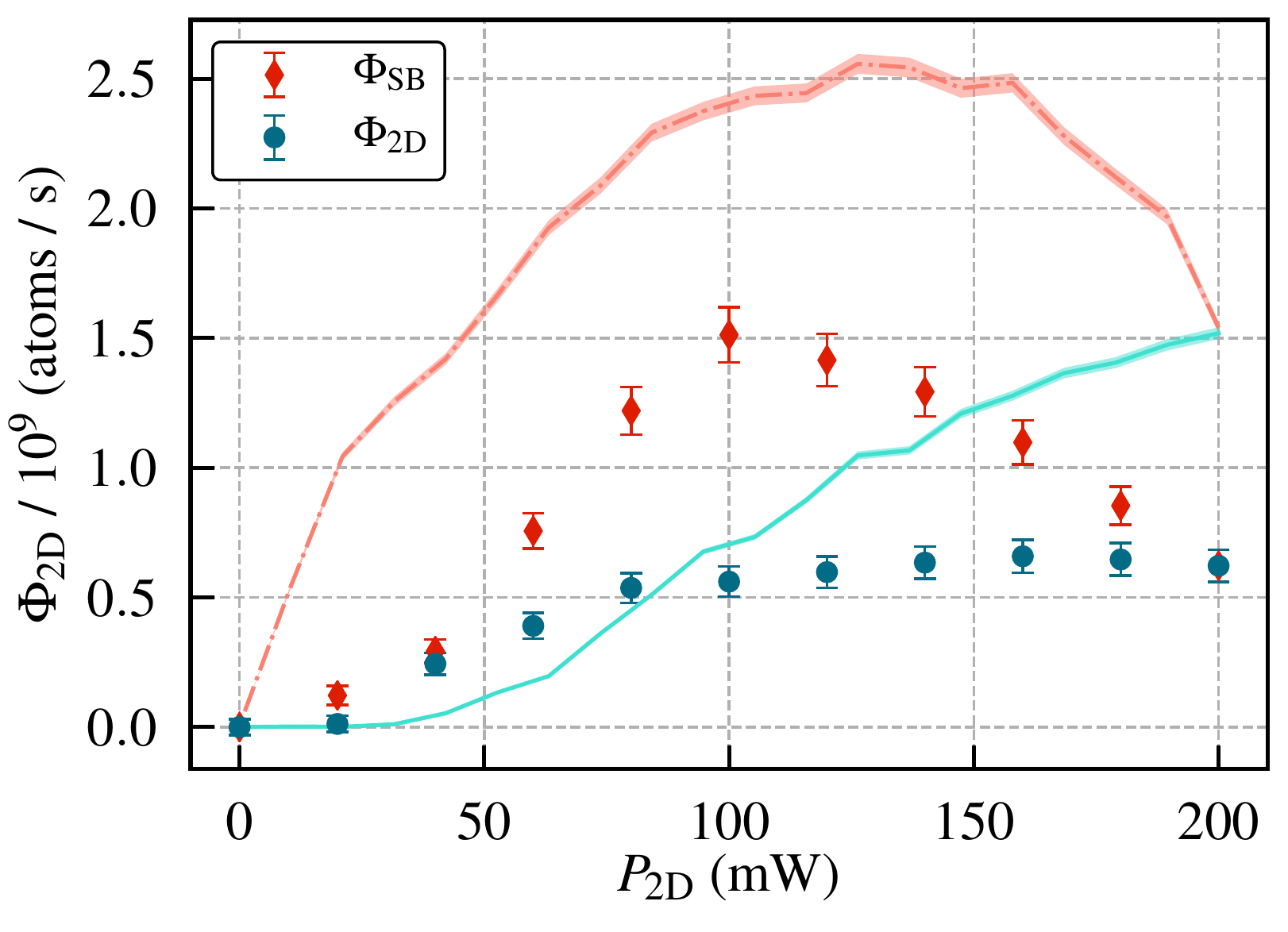}
 \caption{Atomic flux generated from the atomic source as a function of the 2D-MOT saturation parameter, without (blue circles) and with (red diamonds) the use of the sideband ($P_\text{SB}$ = 200 mW - $P_\text{2D}$). These data are taken at $s_\text{push} $ =  0.34,  $\Delta_\text{2D}/\Gamma$ = -1.6 and compared with MC estimates (shaded lines).}
  \label{fig:2Dflux}
\end{figure}

At the best trapping configuration the total atomic flux generated by the 2D-MOT source is measured by detecting the fluorescence generated by a probe beam sent along the $z$-direction, 
nearly at the center of the MOT in the science chamber. The resulting atomic flux $\Phi_\text{2D}$ reaches a maximum value of \SI{6(1)e8}{atoms \per s}, as shown in Fig.~\ref{fig:2Dflux}. This can be compared with the MOT loading rate $L_\text{MOT}$ and with the expected flow resulting from the capture efficiency ratio resulting from MC simulations. The MOT loading rate is simply given by $L_\text{MOT}=N_0/\tau =\SI{3.1(4)e8}{atoms \per s}$, where $\tau$ is the MOT relaxation time, which in our system without repumping is \SI{17(2)}{ms} and $N_0=\SI{5.3(2)e6}{atoms}$ is the maximum number of atoms trapped in the final MOT. It corresponds to roughly \SI{51}{\percent} of the total flux. The expected atomic flux generated by the 2D-MOT can be estimated as 
$$
\Phi^{(th)}_\text{2D} = r f_\text{cut} p_\text{rad} \Phi_\text{ov} = \SI{1.5e9}{atoms \per s}.
$$ 
In this estimate we used $r=\num{6.9e-2}$ from MC results, $f_\text{cut}=\num{1.53e-3}$ is the fraction of simulated velocities from the Maxwell-Boltzmann distribution considering a cut-off at $v_\text{cut}=\SI{90}{m \per s}$, $p_\text{rad}$ is the survival probability from optical pumping to the metastable $^3$P$_2$ state~\cite{Xu2003} that we calculated considering a typical time spent in the 2D-MOT region $\expval{ \tau_\text{2D} }= \SI{2.9(6)}{ms}$ and a pumping rate $R=\SI{223}{Hz}$, which give us $p_\text{rad}\simeq 1 - R \expval{ \tau_\text{2D}} = 0.35$. The estimated theoretical flux $\Phi^{(th)}_\text{2D}$ provides a discrepancy from the measured $\Phi_\text{2D}$ of only a factor 2.4, which is remarkably close. In fact our simulations do not consider effects due to experimental imperfections, such as the misalignment of the zero magnetic field of the permanent magnets and the optimal push beam direction for optical transfer to the MOT in the science chamber.

\section{Sideband enhancement \label{sec:sideband enhancement}}

\subsection{Loading a MOT with sideband-enhancement}

\begin{figure}[b]
    \centering
    \includegraphics[width=0.49\textwidth]{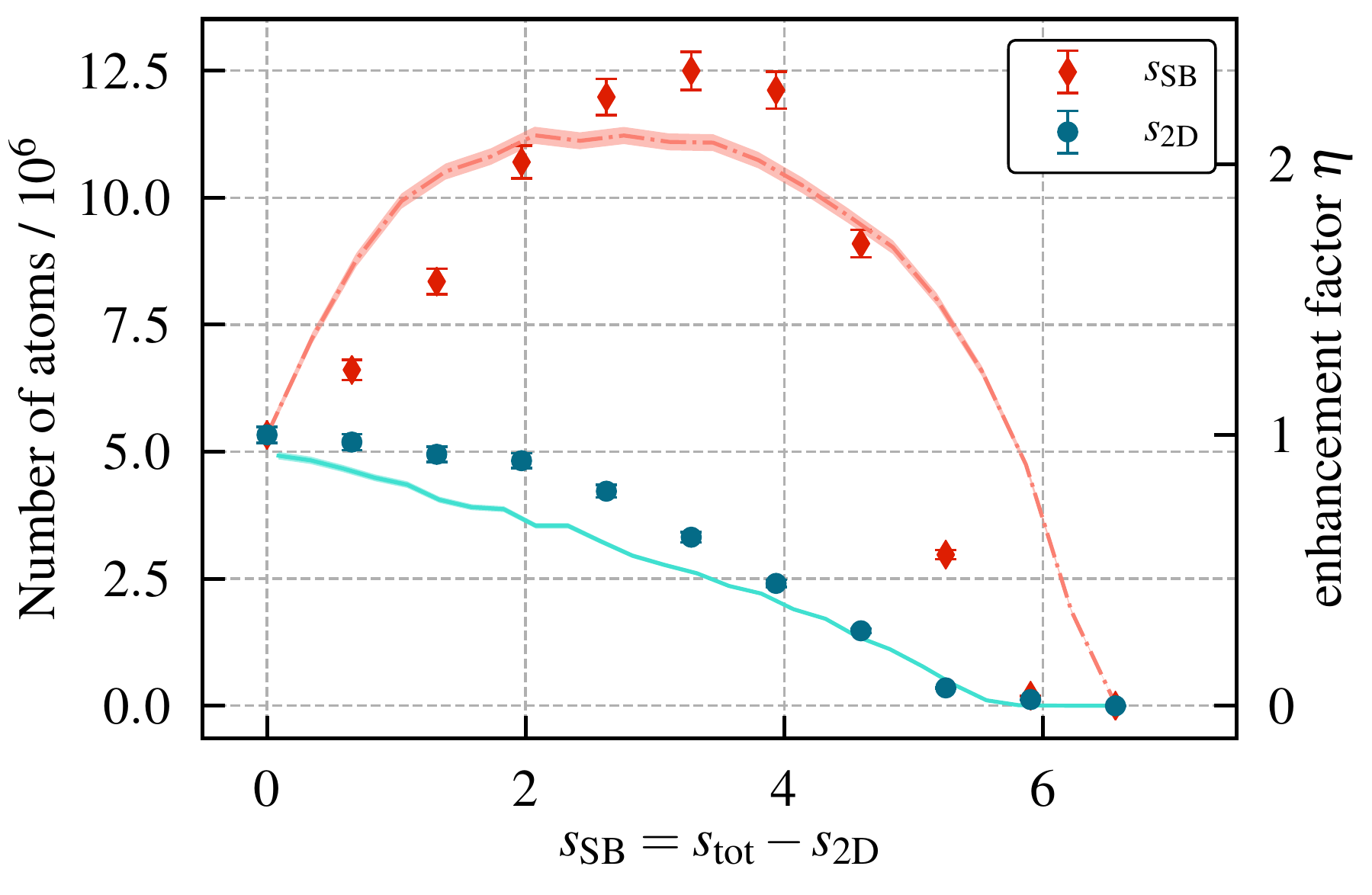}
    \caption{Number of atoms captured in the MOT region for different power distribution of the $s_\text{tot}=6.56$ between the sideband beam $s_\text{SB}$ and the 2D-MOT beam $s_\text{2D}$. All the data are measured with a 2D-MOT detuning $\Delta_\text{2D} = -1.6\,\Gamma$ and sideband detuning $\Delta_\text{SB}= \SI{-3.13}{\Gamma}$. The (red) diamonds are the number of atoms trapped in the MOT at increasing sideband beam saturation parameter $s_\text{SB}$. The (pink) shaded area represents the MC simulation in this experimental configuration. The (blue) circles describe corresponding number of atoms trapped in the MOT with the sideband beam AOM turned off.}
    \label{fig:sideband}
\end{figure}

We demonstrated sideband-enhanced loading of a 2D-MOT atomic source by overlapping a second laser beam with higher frequency detuning to the 2D-MOT cooling lasers, as described in Sec.\ref{sec:apparatus}. Fig.\ref{fig:sideband} shows how the power distribution between the two frequencies affects the number of atoms collected in the MOT trap at sideband detuning $\Delta_\text{SB}=\SI{ -3.13}{\Gamma}$, while the 2D-MOT beam is tuned at its previously shown maximum $\Delta_\text{2D}=\SI{-1.6}{\Gamma}$. From Fig.\ref{fig:sideband} we see that it exists an optimal power distribution around $s_\text{SB} \simeq 3.5$ that maximizes the number of atoms in trapped in the MOT, reaching up to \SI{1.2e7}{atoms} atoms, i.e., about \num{2.3} times higher than with the total available power sent to the 2D-MOT AOM and about \num{4} times higher than the corresponding value with the sideband beam blocked.

In order to find the optimal working point of the sideband-enhanced 2D-MOT, we scanned over the sideband AOM frequency from \SI{230}{MHz} to \SI{335}{MHz}, which corresponds to a detuning range between $-5\Gamma$ and $-2.2\Gamma$, and measured the MOT trapped atoms $N$ at different sideband power. We performed this scan with a total power $P_\text{tot}=\SI{200}{mW}$ and \SI{110}{mW}, i.e. a total saturation parameter $s_\text{tot}=6.56$ and $s_\text{tot}=3.61$, respectively. We introduce the enhancement parameter $\eta$ as: 
\begin{equation}
    \eta(s_\text{SB} , \Delta_\text{SB})= \frac{N(s_\text{SB}= s_\text{tot} - s_\text{2D}  , \Delta_\text{SB})}{ N(s_\text{2D} = s_\text{tot} )  }
    \label{eq:enhacement_factor}
\end{equation}
The so-defined $\eta$ parameter compares the two different trapping configurations, both sharing the same total optical power $s_\text{tot}$. When $\eta > 1$ sideband-enhancement is achieved. 

\begin{figure*}
    \centering
    \includegraphics[width=\textwidth]{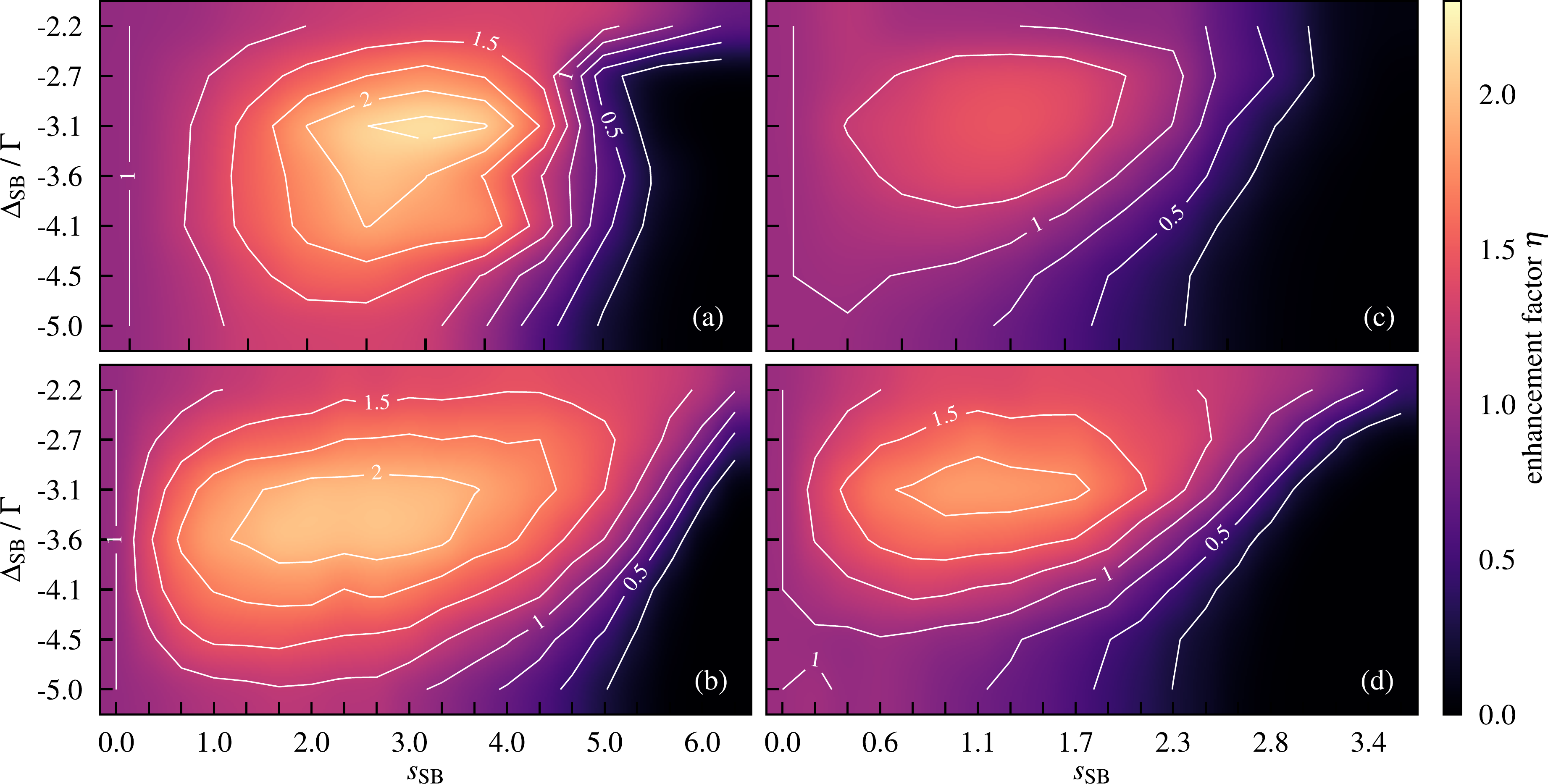}
    \caption{Enhancement factor $\eta$ as function of the sideband parameter scan. (a) Experimental data compared to (b) MC numerical results at fixed total saturation parameter $s_\text{tot}=6.56$ and $\Delta_\text{2D}=-1.6\,\Gamma$. (c) and (d) same as (a) and (b) but for $s_\text{tot}$ = 3.6.}
    \label{fig:SE_scan}
\end{figure*}

Fig.\ref{fig:SE_scan}(a) and (c) show two sets of the sideband enhancement parameter scan, where we plot the enhancement parameter ($\eta$) with respect to $s_\text{SB}$ and $\Delta_\text{SB}$ when $P_\text{tot} = \SI{200}{mW}$ and \SI{110}{mW} respectively. These results are compared to their respective MC simulations (Fig.\ref{fig:SE_scan}(b) and (d)). The data show that optimum loading efficiency of the final MOT is reached tuning the sideband frequency to $\Delta_\text{SB} = \SI{-3.13}{\Gamma}$ for both the total power regimes.  At $P_\text{tot}=\SI{200}{mW}$, we reached a maximum enhancement of $\eta^\text{exp}=2.3(1)$ when $s_\text{SB} \simeq 3.1$ ($P_\text{SB}\simeq \SI{90}{mW}$). The MC numerical data present essentially the same main features as the experimental measurements, both having the maximum loading at the same sideband parameter point, reaching a slightly lower enhancement $\eta^\text{MC} = 2.13(4)$, as detailed in Fig.\ref{fig:sideband}. At $P_\text{tot}=\SI{110}{mW}$, we obtained the best enhancement factor of $\eta^\text{exp}=1.48(7)$ when $s_\text{SB}=1.3$ ($P_\text{SB} \simeq \SI{40}{mW} $), while the MC numerical results show a slightly higher enhancement of $\eta^\text{MC}=1.87(3)$. Both numerical and experimental data suggest that for $s_\text{tot}\geq 1$, sideband-enhancement grows with the increasing available power, where the optimized power distribution can be more effective. Alternatively, the rate at which the atomic flux increases with respect to the laser power is significantly higher for the case where we add the higher-detuned frequency sideband.

At maximum $\eta^\text{exp}$ = 2.3(1), we measured $N=\SI{1.25(4)e7}{atoms}$ trapped in the MOT, which corresponds to a loading rate of $L_\text{MOT}^\text{SB}=\SI{7.3(9)e8}{atoms \per s}$. A total flux measurement by fluorescence detection is also performed for the sideband-enhanced 2D-MOT and reported in Fig.~\ref{fig:2Dflux}. In this case we measured an atomic flux of $\Phi_\text{SB} = \SI{1.5(2)e9}{atoms \per s}$. The related enhancement factor is \num{2.4(4)} and it is in agreement with the experimental and numerical results reported for the MOT loading.

Compared to other Sr atomic sources, our sideband-enhanced 2D-MOT source shows high transfer efficiency $L_\text{MOT}^\text{SB}/\Phi_\text{SB}$ = 48(8)\%, with a MOT loading rate slightly larger than a ZS-enhanced Sr 2D-MOT source~\cite{Nosske2017}, and less than a factor ten lower than more complex and power-demanding high-flux source systems based, for instance, on a combination of Zeeamn slower, 2D-MOT and deflection~\cite{Yang2015}. 


\subsection{Kinetic properties of the sideband-enhanced 2D-MOT}

The kinetic properties of a Sr cold atomic beam generated by a 2D-MOT has been recently studied~\cite{Nosske2017}, in particular as function of the push beam and 2D-MOT beam intensities. We verified these findings in our setup and extended the study to the addition of the sideband beam. 

The longitudinal velocity was measured by time-of-flight technique. A push beam pulse of \SI{5}{ms} accelerate the 2D-MOT atoms towards the science cell. The longitudinal velocity distribution is estimated recording the fluorescence time distribution $f(t)$ 
as measured at the MOT center
We compute the longitudinal velocity distribution as $f(v) = f(d/t)$, where $d = \SI{36.5(5)}{cm}$ is the 2D-MOT to MOT distance. Compared to the single-frequency 2D-MOT, we did not observe any change in peak velocity or in velocity dispersion. The peak velocity for optimal push saturation parameter $s_\text{push} = 0.34$ is $v_\text{L} = \SI{22.5}{m \per s}$.

\begin{figure}[b]
    \centering
    \includegraphics[width=0.49\textwidth]{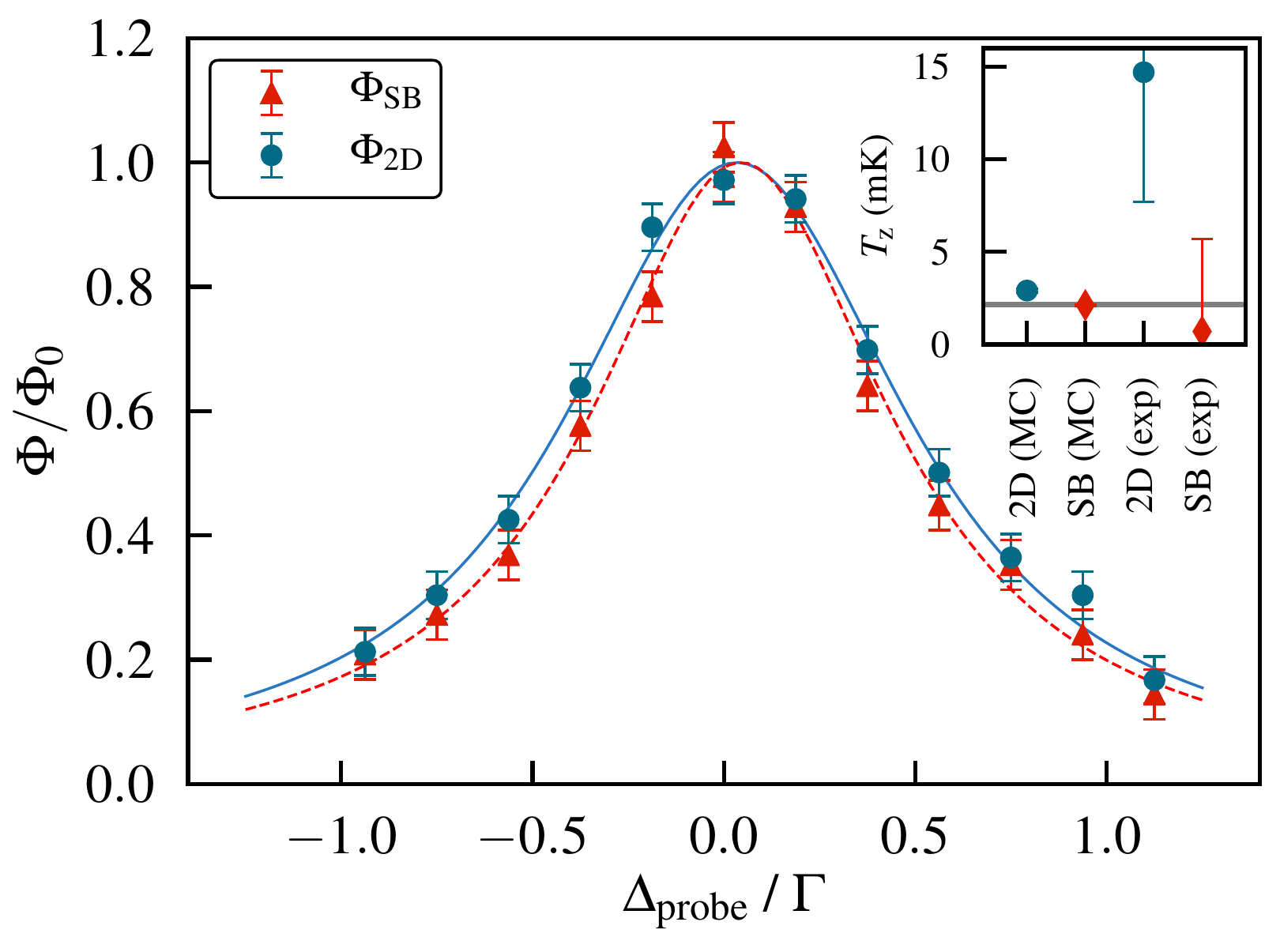}
    \caption{Doppler spectrum of the transverse spectroscopy on the 2D-MOT Sr atomic beam. The inset shows the corresponding Doppler temperatures compared to the MC simulation results and the corresponding Doppler limit.}
    \label{fig:transvel}
\end{figure}

The atomic beam transverse velocity was measured by Doppler spectroscopy with and without powering the sideband beam. The transverse velocity was extracted from the Doppler profile by fixing the Lorentzian component due to the natural linewidth of the \srfc~probe transition and the saturation broadening ($s_\text{probe} = 0.1$). The measurement results are shown in Fig.\ref{fig:transvel}, yielding a Doppler broadening $\sigma_\text{2D}(T) = \SI{3.6(8)}{MHz}$ and $\sigma_\text{SB}(T) = 0.8(3.0)\,\text{MHz}$ respectively. This corresponds to a transverse temperature of \SI{14(7)}{mK} for the 2D-MOT and $0.7(5.0)\,\text{mK}$ for the sideband-enhanced 2D-MOT, as shown in the inset in Fig.\ref{fig:transvel}. Compared to the Doppler temperature at $s_\text{2D} = 6.6$ and $\Delta_\text{2D}=-\SI{1.6}{\Gamma}$ which is equal to \SI{2.1}{mK}, the 2D-MOT result is nearly seven times warmer, while the sideband-enhanced case shows an upper limit temperature almost three times higher. We also estimated the transverse temperature of the atomic beam resulting from MC numerical simulations, which present transverse temperatures of \SI{2.9(1)}{mK} and \SI{2.12(4)}{mK} respectively. While MC results confirms a colder beam for the sideband-enhanced case, they still miss the extra-heating effects which can be explained by transverse spatial intensity fluctuations of the optical molasses in the 2D-MOT~\cite{Chaneliere05}.

From the measured transverse and longitudinal velocities, we derive an atomic beam divergence $\theta_\text{2D} \equiv v_t/v_L = \SI{75(17)}{ mrad}$ and $\theta_\text{SB}\leq \SI{58}{mrad}$. Finally, from the values of the beam divergence and the atomic flux, we estimated the atomic beam radiant intensity, or sometimes called beam ``brightness'', $\mathcal{J}\equiv \Phi /(\pi\theta^2)$. For the single-frequency 2D-MOT we obtained a brightness

$$
\mathcal{J}_\text{2D} = 3.5(8)\times 10^{10} \,\mathrm{atoms}\cdot\mathrm{s}^{-1}\cdot\mathrm{sr}^{-1},
$$
while, for the sideband-enhanced beam, the brightness

$$
\mathcal{J}_\text{SB} \geq 1.4\times 10^{11}\,\mathrm{atoms}\cdot\mathrm{s}^{-1}\cdot\mathrm{sr}^{-1}.
$$

This results represents a factor four improvement with respect to the single-frequency 2D-MOT, making the sideband-enhancement a promising technique for optimal transfer to 2D optical molasses working on narrow linewidth intercombination transition of strontium for continuous BEC production and continuous optical clock proposals~\cite{Bennetts17}.

\subsection{Comparison with the Zeeman-slower enhancement}

An alternative method to increase the 2D-MOT capture rate is to direct another slowing beam towards the hot atomic beam generated by oven which, exploiting the decreasing tail of the 2D-MOT magnetic field, can efficiently scatter faster atoms along the beam direction similarly to a Zeeman Slower (ZS). This approach was previously demonstrated in similar setups~\cite{Lamporesi2013,Nosske2018,Colzi2018}.

We employed the beam generated by the sideband AOM as Zeeman slowing beam, shaped to have a beam width $w_{ZS} = \SI{6}{mm}$. We partially scanned over the ZS parameters, which resulted in a maximum number of atoms in the MOT of $N_\text{ZS} = \SI{3.8(1)e6}{atoms}$, obtained with $\Delta_\text{ZS}=\SI{-8.1}{\Gamma}$, $P_\text{ZS} =\SI{140}{mW}$, while we kept $P_\text{2D} =\SI{36}{mW}$ and $\Delta_\text{2D}=\SI{-1.5}{\Gamma}$ fixed. Blocking the ZS beam we observe a gain in the atomic number of the order of 4, in agreement with the experimental observation in~\cite{Nosske2017}. 

Because of the short distance (about \SI{25}{cm}) between the oven aperture and the optical window facing it to send the ZS beam, we heated the window flange up to \SI{350}{\celsius} in order to prevent metalization. However we first observed a fast degradation of the atomic source flux, soon followed by the full metalization of the window. This prevented us to perform a fine optimization of the $s_\text{ZS}$ and $\Delta_\text{ZS}$ as the one shown in the Fig.~\ref{fig:SE_scan} and also to produce a stable MOT during the day. 

A possible way to reduce Sr metalization of the ZS window would be to increase the distance between the oven and window itself by means of a vacuum extension (at least a \SI{1}{m} tube with shorter diameter). However this solution would compromise the compactness of the atomic source conceived by a 2D-MOT making the system more complex, power consuming and perhaps needing extra water cooling to avoid thermal stress to the vacuum system.

Another drawback of the ZS method is the fact that the quantization axis of the magnetic field along the hot atoms direction imposes that only one half of the linear polarized ZS light power has the correct circular polarization. This means that at least half of the ZS optical power is wasted in the slowing process. On the contrary, the sideband beams have a well defined polarization in the capture region so that all the employed power is effective in the cooling and trapping process. 

\subsection{Application to other alkaline-earth atoms and prospects for optical clocks}

It is interesting to extend the discussion about the sideband-enhancement method to other atomic species, in particular those employed in optical clocks. We exploit the MC simulation 
in order to investigate the potential trapping performances of additional atomic species. Table~\ref{tab:alkaline-earth} shows the main optical and atomic parameters for alkaline-earth(-like) atomic species currently in use in optical clock experiments. In particular, we consider the broad \srfc~strong dipole transition as cooling transition, fixing the atomic vapour pressure to \SI{0.1}{Pa} for all the species. We ran our MC simulation with these parameters, with a total available saturation parameter $s_\text{tot}$ = 6.56, the same magnetic field gradient and the same laser beam widths as for our previously described apparatus. 

The simulation workflow is the following: first we simulate the single-frequency 2D-MOT, looking for the optimal detuning $\Delta_\text{2D}$ at half of $s_\text{tot}$; then we add the sideband at $\Delta_\text{SB} = 2\Delta_\text{2D}$, which is basically the result we found in Sec.\ref{sec:sideband enhancement} for Sr, and we scan the sideband-enhanced 2D-MOT at different sideband saturation parameter $s_\text{SB}$. Because of the extremely high saturation intensities of Cd and Hg which makes unrealistic the application of this method, they were excluded from this numerical study.

\begin{table*}[tb]
\centering
\begin{tabular}{l c c c c c c |c c c}
\hline
\hline
Atom & $\lambda$     &    $\Gamma/2\pi$  & $I_\text{sat}$ & $a_\text{max}$ & $T(p_0)$& $v_{th}(p_0)$ & $\Delta_\text{2D}/\Gamma$ & $r^\text{MC}f_\text{cut}$ & $\eta$\\
 & (nm) & (MHz) & (mW/cm$^2$) & ($10^6$ m/s$^2$) & (K) & (m/s)&&(ppm)\\
\hline
$^{24}$Mg  & 285.30 & 80.95 & 455 & 14.8& 641 & 679 & -2.28& 77 & 1.0\\
$^{40}$Ca  & 422.79 & 34.63 & 59.9& 2.69 & 788 & 583 & -2.5 & 103& 2.1\\
$^{88}$Sr  & 460.86 & 31.99 & 42.7 & 1.01 & 725 & 379 &-1.76& 260& 2.1\\
$^{138}$Ba  & 553.70 & 18.33 & 14.1& 0.31 & 826 & 321&-0.89& 60&2.9 \\
$^{114}$Cd  & 228 & 91& 1005 & 4.64 & 485 & 273\\
$^{174}$Yb  & 398.91 & 29 & 59.8 & 0.54 & 673 & 258&-1.42& 316& 2.2\\
$^{198}$Hg & 185 & 120 & 2481 & 4.24 & 286 & 157\\

\hline
\hline
\end{tabular}
\caption{Sideband enhancement on alkaline-earth(-like) atomic species. On the left side of the table the most relevant parameters for cooling and trapping atoms on the \srfc~ strong dipole transition are shown, whereas we estimate the thermal and kinetic properties of every atomic species at a pressure $p_0$ = 0.1 Pa. On the right, MC optimization results of the 2D-MOT detuning $\Delta_\text{2D}$, the fraction of trapped atoms, and the enhancement factor $\eta$ for each alkaline-earth species.} 
\label{tab:alkaline-earth}
\end{table*}

The MC simulation results are reported in Tab.~\ref{tab:alkaline-earth}. Here we clearly see that the sideband-enhancement is more effective for those atoms having a lower value of the maximum acceleration $a_\text{max}$. This dependence can be understood by looking at the definition of maximum capture velocity in (\ref{eq:vc}). In fact, it can be only achieved for a light field uniformly resonant with the atomic transition and fully saturated all along the trap diameter. This means that the broader the cooling transition linewidth $\Gamma$ is (and thus the higher $a_{max}$), the closer the MOT is to its capture limit, implying that the expected enhancement factor is lower. Furthermore, according to Eq.~\ref{eq:cap_2d}, we would expect that the sideband-enhancement works better for light species, in particular where $v_\text{th}$ is higher and the quartic dependence of the loading rate on the capture velocity is a more accurate approximation. Hence we can work out a sideband-enhancement factor functional dependence

$$
\eta(X) \propto \frac{v_\text{th}(X)^2}{a_\text{max}(X)}
$$
where $X$ is the considered atomic species. By accident, either Sr, Ca and Yb have very similar ${v_\text{th}(X)^2}/{a_\text{max}(X)}$ values, and the resulting $\eta$ is the same within the numerical error.

We also report in Tab.\ref{tab:alkaline-earth} the absolute capture efficiency for the sideband-enhanced 2D-MOT atomic source $r(X)$. MC simulations show that the highest capture rate is predicted for Yb followed by Sr, two of the strongest candidates for a possible redefinition of the second based on optical atomic clocks~\cite{Riehle2015}.


\section{Conclusions}\label{sec:conclusion}

In this work we demonstrated and fully characterized a robust method to enhance the atomic flux generated by a Sr 2D-MOT by adding a second frequency to the 2D-MOT beams. The experimental implementation of the sideband-enhancement method only requires a simple optical setup and a proper alignment of the sideband beam to the main 2D-MOT beam. The resulting bright atomic source can deliver more than $1.4\times 10^{11}\,\mathrm{atoms}\cdot\mathrm{s}^{-1}\cdot\mathrm{sr}^{-1}$ if the total available power for the atomic source is 200 mW. This cold atomic flux can be efficiently loaded in a 3D MOT for ultracold atoms experiments, preventing direct sight to the hot atomic oven and providing an efficient optical shutter of the atomic beam. This result represents an enhancement in MOT loading by a factor 2.3 with respect to single-frequency 2D-MOT based atomic source.

A dedicated Monte Carlo simulation, which well predicts the experimental data of our Sr atomic source, shows that this technique is a valid method to increase the number of atomic sources based on the other alkaline-earth species such as Yb and Ca, paving the way for compact atomic sources suitable for transportable optical clocks or optical clock transition-based gravimeters~\cite{Hu17,Akatsuka_2017}. 



\begin{acknowledgments}
The authors would like to thank U. Sterr for inspirational discussions about sideband-enhanced MOT, D. Racca, E. Bertacco, M. Bertinetti and A. Barbone for laboratory assistance. We acknowledge funding of the project EMPIR-USOQS, EMPIR projects are co-funded by the European Union’s Horizon2020 research and innovation programme and the EMPIR Participating States.
We also acknowledge QuantERA project Q-Clocks, ASI, and Provincia Autonoma di Trento (PAT) for financial support. 
\end{acknowledgments}

%


\end{document}